\documentclass[a4paper,11pt]{article}
\usepackage{amsmath}
\usepackage{color}
\usepackage{graphicx}
\usepackage{amssymb}
\usepackage{float}
\usepackage{scrtime}
\usepackage{fancyhdr}
\usepackage{subfigure}
\usepackage{authblk}
\usepackage{lipsum}
\usepackage{rotating}
\usepackage{floatpag}
\usepackage{varioref}
\usepackage{slashed}
\usepackage{enumerate}
\usepackage{relsize}
\usepackage[toc,page]{appendix}

\usepackage[colorlinks=true
,urlcolor=blue
,anchorcolor=blue
,citecolor=blue
,filecolor=blue
,linkcolor=black
,menucolor=blue
,pagecolor=blue
,linktocpage=true
]{hyperref}

\usepackage[numbers,sort&compress]{natbib}
\usepackage{cancel}

\setlipsumdefault{131}

\newcommand{\ibar}{\overline{i}}
\newcommand{\jbar}{\overline{j}}

\newcommand{\lbar}{\overline{l}}

\newcommand{\nbar}{\overline{n}}

\newcommand{\Fbar}{\overline{F}}
\newcommand{\Qbar}{\overline{Q}}

\newcommand{\be}{\begin{equation}}
\newcommand{\ee}{\end{equation}}
\newcommand{\beq}{\begin{equation}}
\newcommand{\eeq}{\end{equation}}
\newcommand{\bea}{\begin{eqnarray}}
\newcommand{\eea}{\end{eqnarray}}
\newcommand{\bei}{\begin{itemize}}
\newcommand{\eei}{\end{itemize}}

\newcommand{\gsim}{\lower.7ex\hbox{$\;\stackrel{\textstyle>}{\sim}\;$}}
\newcommand{\lsim}{\lower.7ex\hbox{$\;\stackrel{\textstyle<}{\sim}\;$}}

\newcommand{\ba}{\begin{array}}
\newcommand{\ea}{\end{array}}
\newcommand{\et}{\end{tabular}}
\newcommand{\bc}{\begin{center}}
\newcommand{\ec}{\end{center}}

\def\vev#1{\mathopen\langle #1\mathclose\rangle }
\def\nn{\nonumber}

\def\0 {\nonumber} 
\def\del{\partial} 

\def\Ox{\Omega} 
\def\bj{\bar \jmath}

\newcommand{\Kh}{{\hat{K}}}

\newcommand{\sym}[2]
   {\omega\!\left(#1,#2\right)}
\newcommand{\symr}[2]
   {\omega_\rho\!\left(#1,#2\right)}
\newcommand{\symJ}[2]
   {\omega_J\!\left(#1,#2\right)}

\newcommand{\cref}{{\bf [check ref]}}

\newcommand{\Y}{{\hat Y}}

\newcommand{\N}{\mathcal{N}}

\newcommand{\Wh}{\hat{W}}

\labelformat{subsection}{Section #1} 
\labelformat{equation}{Eq.~(#1)} 
\labelformat{figure}{Fig.~#1} 
\labelformat{subfigure}{Fig.~\thefigure#1} 
\labelformat{table}{Tab.~#1}

\def\be{\begin{equation}}
\def\ee{\end{equation}}
\def\bea{\begin{eqnarray}}
\def\eea{\end{eqnarray}}
\def\nn{\nonumber}

\graphicspath{{./figures/}}

\begin{document}

\floatpagestyle{plain}

\pagenumbering{roman}

\renewcommand{\headrulewidth}{0pt}
\fancyfoot{}

\title{\huge \bf{Early Universe Cosmology, Effective Supergravity, and Invariants of Algebraic Forms}}

\author{Kuver Sinha}

\affil{\em Department of Physics, Syracuse University, 
Syracuse, NY 13244, USA\enspace\enspace\enspace\enspace\enspace}

\maketitle

\thispagestyle{fancy}


\begin{abstract}\normalsize\parindent 0pt\parskip 5pt

The presence of light scalars can have profound effects on early universe cosmology, influencing its thermal history as well as paradigms like inflation and baryogenesis. Effective supergravity provides a framework to make quantifiable, model-independent studies of these effects. The Riemannian curvature of the Kahler manifold spanned by scalars belonging to chiral superfields, evaluated along supersymmetry breaking directions, provides an order parameter (in the sense that it must necessarily take certain values) for phenomena as diverse as slow roll modular inflation, non-thermal cosmological histories, and the viability of Affleck-Dine baryogenesis. Within certain classes of UV completions, the order parameter for theories with $n$ scalar moduli is conjectured to be related to invariants of $n$-ary cubic forms (for example, for models with three moduli, the order parameter is given by a function on the ring of invariants spanned by the Aronhold invariants). Within these completions, and under the caveats spelled out, this may provide an avenue to obtain necessary conditions for the above phenomena that are in principle calculable given nothing but the intersection numbers of a Calabi-Yau compactification geometry. As an additional result, abstract relations between holomorphic sectional and bisectional curvatures are utilized to constrain Affleck-Dine baryogenesis on a wide class of Kahler geometries.


\end{abstract}

\clearpage

\newpage
\pagenumbering{arabic}

\section{Introduction} \label{intro}

$D=4$, $\N=1$ effective supergravity is the setting for much of phenomenology and cosmology, especially work that is "string-inspired", with varying degrees of UV completions \cite{McAllister:2007bg}. In this paper, we will be interested in three cosmological applications: slow roll moduli inflation \cite{Baumann:2014nda, Burgess:2011fa, Baumann:2009ni, Cicoli:2011zz}, the thermal history of the Universe prior to Big Bang Nucleosynthesis \cite{Kane:2015jia}, and Affleck-Dine Baryogenesis \cite{Dine:1995uk, Dine:1995kz, Affleck:1984fy}.

It is remarkable that such a wide range of cosmological phenomena can be described within the same setting. This is because of several key factors that can play an important role in early universe cosmology and are naturally captured within a supergravity framework:

\begin{itemize}

\item The importance of moduli: Due to their gravitational coupling to other fields, moduli  can play an important role in the thermal history of the Universe if they come to dominate the energy density \cite{Moroi:1999zb, Dutta:2009uf, Acharya:2008bk}. Moreover, moduli potentials are particularly suited for small-field slow roll inflationary models \cite{BlancoPillado:2004ns, Linde:2007jn, Cicoli:2008gp, Allahverdi:2009rm}, due to their flatness in perturbation theory. 

\item Supersymmetric flat directions: Supersymmetric flat directions play an important role in early universe cosmology. By stabilizing scalar potentials against quantum corrections, supersymmetry helps to satisfy flatness conditions required for slow roll inflation\footnote{However, to solve the $\eta$ problem in supergravity requires extra fine-tuning.}. On the other hand, supersymmetric flat directions in the visible sector participate in robust frameworks like the Affleck-Dine mechanism that can explain the matter- anti-matter asymmetry of the Universe.

\item Inflation breaks supersymmetry: The vacuum energy during inflation breaks supersymmetry, leading to soft terms ("Hubble-induced terms") in the visible sector Lagrangian.

\end{itemize} 

Within supergravity, the scalar potential $V$ should allow for spontaneous supersymmetry breaking with the following features:

\begin{itemize}

\item Phenomenology: Acceptable phenomenology requires a point in moduli space where $V \, \sim \, 0$, $V^{\prime} \, = \, 0$, and $V^{\prime \prime} \, > \, 0$ are necessarily true. The first is the requirement of vanishing cosmological constant in the present Universe, the second is the requirement that the vacuum energy is extremized, and the third is the requirement that we live in a (meta)stable Universe.

\item Cosmology (Inflation): To obtain a viable period of slow roll modular inflation, the scalar potential necessarily needs to satisfy   $V \, \sim \, H^2$, $V^{\prime} \, \sim \, 0$, and $V^{\prime \prime} \, \lsim \, 0$ at some point in field space. The first condition is the requirement of inflationary vacuum energy, while the second and third are required to ensure the smallness of the slow-roll parameters $\epsilon$ and $\eta$. We reiterate that our interest in this paper will be entirely slow roll modular inflation; in particular, our treatment does not apply to axionic inflation models, brane - anti-brane models, visible sector models etc.


\item Cosmology (Thermal History): Moduli can dominate the energy density of the Universe and, being gravitationally coupled, can decay late. When the lifetime exceeds the onset of Big Bang Nucleosynthesis, entropy dilution from the decay ruins the successful prediction of the abundances of the light elements; this is the cosmological moduli problem\footnote{Originally the Polonyi problem, dating from the earliest theories of supergravity \cite{Coughlan:1983ci, Hall:1983iz, Banks:1993en}.}. The lifetime depends on the modulus mass; hence thermal history is determined by points in moduli space that satisfy $V \, \sim \, 0$, $V^{\prime} \, \sim \, 0$, and $V^{\prime \prime} \, \sim \, m^2_{3/2}$. The first two conditions are the cosmological constant and stability criteria, while the last is a statement that the modulus decays around the time of BBN, assuming that supersymmetry is broken at low energies with $m^2_{3/2} \sim \mathcal{O}(100)$ TeV.

\item Cosmology (Affleck-Dine Baryogenesis): Unlike other baryogenesis mechanisms, effective supergravity is the natural setting for Affleck-Dine baryogenesis \cite{Dutta:2010sg, Garcia:2013bha, Dutta:2012mw, Marsh:2011ud}. A visible sector baryon number carrying flat direction acquires a tachyonic mass during inflation, rolls to non-zero vacuum expectation value, and acquires a CP-violating decay from the competing effects of Hubble-induced and soft A-terms. In terms of supergravity data, the necessary requirement is that at some point in field space the conditions $V \, \sim \, H^2$, and $V^{\prime \prime}_{vis} \, < \, 0$ hold. The first condition corresponds to the inflationary vacuum energy, while the second condition is the requirement that the Hubble-induced mass on some visible sector flat direction is tachyonic.

\end{itemize}

The full data of a string compactification, including the locations of D-branes, orientifolds, flux quanta, anomaly cancellation, non-local effects on the potential, and local visible sector model building would of course settle the question of whether these conditions can be satisfied in a UV complete setting. That is a challenging task (for a survey of these topics, we refer to the excellent reviews listed above).

Since the scenarios listed above depend crucialy on certain local analytic properties of the potential $V$ of the scalar components of chiral superfields, it is possible to ask questions purely at the level of local geometry, without referring to a particular UV completion, or even the full global details of the potential.
 %
%
For generic directions in chiral multiplet space, the values of $V, \, V^{\prime},$ and $V^{\prime \prime}$ depend both on the Kahler potential $K$ and superpotential $W$, and saying anything predictive amounts to knowing both quantities, i.e., the full data  that (along with the gauge kinetic function) determines the supergravity Lagrangian. However, as long as one is satisfied in making \textit{necessary} but not \textit{sufficient} statements, it is indeed possible to obtain a simple, local, geometric order parameter (we will use the term "order parameter" to denote some quantity that must \textit{necessarily} take certain values in order for a phenomenon to take place). A study of the scalar potential shows that this parameter is the local curvature \cite{Covi:2008ea, Covi:2008cn, Covi:2008zu}. Specifically, it is the component of the  Riemann curvature tensor along the Goldstino directions defined by the auxiliary fields that encode the supersymmetry breaking data
\be \label{basicntro1}
{\rm order \,\, parameter:\,\,} R[f] \,\,\,\,\, \equiv \,\,\,\, \, R_{i \jbar m \nbar} f^i f^{\jbar} f^m f^{\nbar} \,\,\,\,\, \equiv \,\,\,\,\, \mathlarger{\mathlarger{\mathlarger{\sum}}}_{\substack{i,j,k \,\equiv \\ {\rm all} \,\, \cancel{SUSY} \\ {\rm directions} }} \frac{R_{i \jbar k \lbar}}{g_{i \jbar} g_{k \lbar}} \,\,.
\ee
The necessary conditions on this quantity are listed in \ref{HoloBiholoresults}. Here, $f^i \, \equiv \, F^i/|F|$ is a chiral field with non-vanishing F-term. The important point is that these bounds on the Riemannian curvature have to be satisfied regardless of the details of the UV completion. 
%
%
Moreover, since $R[f]$ is a function of the scalar fields, the value of the quantity in \ref{basicntro1} is implicitly dependent on the superpotential $W$, which (along with $K$) determines the allowed field values in the effective theory.

One may ask if it is possible to recast the conditions on the parameter $R[f]$ in terms of equivalent conditions on other parameters (which we call $\Delta$ for now) that are entirely field-independent, constructed from for example the input parameters that specify the Kahler geometry.  The reason this is useful is the following. If a quantity $\Delta$ serves as an order parameter, then in principle it is possible to rule out certain cosmological phenomena on classes of manifolds $\mathcal{M}$ based solely on the value of $\Delta(\mathcal{M})$, which is given by the geometric input that specifies the Kahler potential of $\mathcal{M}$, \textit{without} any other physical ingredient at all.

As an example, one can consider the case of Calabi-Yau compactifications, where these input parameters are the intersection numbers. The Kahler potential of Kahler moduli in type IIB Calabi-Yau compactifications is given by
\be
K \, = \, -2 \ln{\mathcal{V}} \,\,,
\ee
where $\mathcal{V}$ is the volume form, given in terms of intersection numbers in a basis of two-cycles. As a first guess to a plausible $\Delta$, we can require that the conditions on $\Delta$ should not change under a change of the basis of two-cycles in terms of which the geometric data is presented. This means that $\Delta$ should be a function of the ring of covariants of the volume form $\mathcal{V}$. If further one requires the conditions on $\Delta$ to be field-independent, then $\Delta$ should be a function of the ring of \textit{invariants} of $\mathcal{V}$. We note that invariants of a polynomial $\mathcal{V}$ are polynomials in the coefficients (and \textit{only} the coefficients) of the polynomial $\mathcal{V}$ which remain invariant under linear transformation of the variables. For example, the invariants of the Calabi-Yau volume form are polynomials in its intersection numbers whose forms do not change under the rotation of basis divisors.  Given a database of Calabi-Yau geometries specified by intersection numbers, one requires no other data in the effective supergravity theory (the matter Kahler metric or the superpotential) to check whether necessary conditions for the cosmological pheomena are satisfied on a given manifold. This makes such quantities suitable for computational studies along the lines of \cite{Altman:2014bfa, He:2013epn, Gray:2012jy, Kreuzer:2002uu}.


In the specific supergravity theories taken in Section \ref{invforms1}, we show that this is indeed possible, under several restrictive conditions. If the UV completion is type IIB string theory compactified on a Calabi-Yau, it can be shown explicitly in the case of compactifications with two moduli (neglecting $\alpha^{\prime}$ and $g_s$ corrections) that the order parameter is an invariant (the discriminant) of the Calabi-Yau volume form. Generalising to the case of $n$ moduli, we conjecture that the relevant order parameter (again, neglecting $\alpha^{\prime}$ and $g_s$ corrections) is the discriminant of a $n-$ary cubic, generated by its ring of invariants. The discriminant for an arbitrary $n$ moduli compactification is a polynomial of degree $n.(2)^{n-1}$.  For example, for the case of three moduli, the ring of invariants is generated by the two Aronhold invariants $S$ and $T$, of degree 4 and 6 respectively, and the discriminant is a homogeneous polynomial of degree 12 in the intersection numbers, given by $\Delta_3 = S^3 - T^2$.
%
%
\be
{\rm order \,\, parameter:\,\,discriminant \,\,}\Delta_n \,\,.
\ee

We use these results to study inflation and moduli dynamics. In the simplest case of two moduli, we are able to make definitive statements regarding the possibility of slow roll modular inflation depending on the sign of $\Delta_2$. These results are given in \ref{delta2ordparam}. The case for $n$ moduli is substantially more involved; nevertheless, we conjecture that it is the sign of $\Delta_n$ that is important, and give partial results towards this conjecture in \ref{discrimnmodcase}. Similarly, for the two moduli case, we prove that moduli masses are bounded by the gravitino mass in \ref{modboundfor2modcomp}, while giving indications that a similar result should hold in the general $n$ moduli case as well.

%
%

Apart from the above considerations, we also present a different and complementary method of extracting information at the level of the geometry. At a given point in moduli space, the quantity in \ref{basicntro1} is the sum over holomorphic sectional curvatures of planes in tangent space that are spanned by supersymmetry breaking directions. On the other hand, the induced soft mass ($V^{\prime \prime}_{vis}$) along a visible sector field $Q^\alpha$ depends on the quantity
\be \label{basicntro2}
R[f,Q] \,\,\,\,\, \equiv \,\,\,\, \, R_{i \jbar Q_{\alpha} \Qbar_{\overline{\beta}}} f^i f^{\jbar} Q^\alpha \Qbar^{\overline{\beta}} \,\,\,\,\, \equiv \,\,\,\,\, \mathlarger{\mathlarger{\mathlarger{\sum}}}_{\substack{i,j \,\equiv \, \cancel{SUSY} \\  \alpha, \beta \, \equiv \, {\rm vis}}} \frac{R_{i \jbar \alpha \overline{\beta}}}{g_{i \jbar} g_{\alpha \overline{\beta}}} \,\,.
\ee
At a given point in the space of moduli and visible sector fields, this is the holomorphic \textit{bisectional} curvature of the planes in tangent space spanned by supersymmetry breaking moduli and visible sector fields $Q$ \cite{GoldbergKob}. 

In Section \ref{holobiholosec1}, we take the point of view that by exploring abstract relations between holomorphic sectional (\ref{basicntro1}) and bisectional (\ref{basicntro2}) curvatures, one can constrain Hubble induced soft masses in a model-independent way on classes of manifolds. As an application, we consider the case of Affleck-Dine baryogenesis, which requires Hubble-induced soft masses to be tachyonic ($V^{\prime \prime}_{vis} \, < \, 0$). We find a no-go result:  Affleck-Dine baryogenesis and modular inflation are incompatible on complex space forms, which are Kahler manifolds with isotropic holomorphic sectional curvature at every point. This generalizes an easily verifiable result for the case of symmetric coset manifolds.

The rest of the paper is structured as follows. In Section \ref{effsugra}, we review $D=4, \, \N=1$ supergravity and set our notation. In Section \ref{cosmo1}, we describe the various cosmological phenomena we are interested in and the relevant bounds on the sectional curvature $R[f]$. In Section \ref{invforms1}, we describe the connection to algebraic invariants. In Section \ref{holobiholosec1}, we give the relations between holomorphic sectional and bisectional curvatures. We end with our Conclusions. Most of the calculations are relegated to several Appendices.

\section{Mass Relations in Effective Supergravity} \label{effsugra}

The setting for much of our work will be $\N=1$, $D=4$ effective supergravity whose main features we briefly review in this Section, following \cite{Kaplunovsky:1993rd}.

We will in general be interested in supergravities that have an observable or visible sector (which will be a supersymmetric extension of the Standard Model) and a modulus sector. We will generally leave the visible sector unspecified; depending on the UV completion, different extensions of the Minimal Supersymmetric Standard Model (MSSM) can be constructed. The chiral superfields in the visible sector will be labelled by $Q^I$, and will include all the quark, lepton and Higgs superfields of the MSSM, and possibly additional particles (generally with $\mathcal{O}(1)$ TeV masses).

As or the moduli or 'hidden' sector, we will denote the chiral superfields by  $\Phi^i$. Their vacuum expectation values (vevs) $\vev{\Phi^i}$
    parametrize continuous families of the string vacua. While we will consider a UV completion later, right now it is enough to keep in mind that the effective potential of the moduli can receive both perturbative and non-perturbative contributions, a combination of which will stabilize them to finite vevs. These contributions can also lead to spontaneous supersymmetry breaking in the modulus sector, signalled by a non-zero vev of an $F$-term or a $D$-term. These vevs are the auxiliary components of chiral and vector superfields.

We thus have the following assumptions about the moduli sector:

$(i)$ $V_{eff}(\Phi)$ has a stable minimum, without flat directions.

$(ii)$ At that minimum, $V_{eff}(\vev\Phi)=0$, that is,  the effective cosmological constant vanishes. Of course, this does not "solve" the cosmological constant problem, but rather reflects a fine-tuning of $V_{eff}(\vev\Phi)$. 

$(iii)$ Some of the $\vev{F^i}$  in the moduli direction are non-zero.

The Lagrangian of the effective  supergravity theory is given in terms of gauge couplings (that are moduli dependent in ways that depend on the specific UV completion), the Kahler function $K$ (gauge-invariant real analytic
function of the chiral superfields) and the
superpotential $W$ (holomorphic function of the chiral superfields).

We begin with the superpotential for the effective theory of the moduli $\Phi^i$ and the observable chiral superfields $Q^{\alpha}$, which generally looks like  $W_{full} = \Wh (\Phi) + W_{matter}$, where
\be \label{wmatt1}
W_{matter}(Q^{\alpha})\ = \ \frac{1}{2} \mu_{\alpha \beta}Q^{\alpha}Q^{\beta} \
+\ \frac{1}{3} Y_{\alpha \beta \gamma} Q^{\alpha} Q^{\alpha} Q^{\alpha} ,
\ee
is the classical superpotential. The modulus superpotential can be written schematically as
\be
\Wh (\Phi) \ = \ W_{tree} \ + \ W_{n-p} \,\,,
\ee
where $W_{n-p}$ stands for possible non-perturbative corrections to the superpotential that are crucial to obtaining stabilized moduli. The superpotential does not suffer from renormalization in any order of  perturbation theory.

The Kahler function $K$  is responsible for the kinetic terms in the Lagrangian. Expanding  in powers of $Q^{\alpha}$ and $Q^{\overline{\alpha}}$, we have
\be
K_{full} \ = \ \Kh (\Phi,\overline{\Phi}) \
+\ Z_{\overline{\alpha} \beta}(\Phi,\overline{\Phi})\, \overline{Q}^{\overline{\alpha}}Q^{\beta}  \ + \ \left(\frac{1}{2}
H_{\alpha \beta}(\Phi,\overline{\Phi})\, Q^{\alpha}Q^{\beta} \ + \ {\rm H.c.} \right)
+\ \cdots,
\ee
where the $\cdots$
stand for the higher-order terms;
$Z_{\overline{\alpha} \beta}$ is the Kahler metric for the observable superfields;
the Kahler metric for the moduli is given by
$K_{\overline{i} j} \equiv \bar\partial_{\overline{i}} \partial_j K$. In general, neither the higher-order terms nor the $Z_{\overline{\alpha} \beta}$ are calculable in a model-independent manner.


The effective potential for the moduli, which will be very important for us, is given by
\be \label{effV1}
V_{eff}(\Phi,\overline{\Phi})\ = \ \Kh_{i\jbar}F^i\overline{F}^{\jbar} \
- \ 3 e^{\Kh} |\Wh (\Phi)|^2 ,
\ee
where
\be
\overline{F}^{\jbar} \ = \ e^{\Kh /2}\,
\Kh^{\jbar i} \left( \del_i\Wh + \Wh \del_i\Kh \right) \,,\qquad
\Kh^{\jbar i}\,=\,(\Kh_{i\jbar})^{-1}.
\ee
At the minimum of \ref{effV1}, $V_{eff}(\Phi,\overline{\Phi}) = 0$,
but (some) $\vev{F^i} \neq 0 $ and thus supersymmetry
is spontaneously broken. The measure of this breakdown is the gravitino mass
\be
m_{3/2} \ = \ e^{\vev\Kh /2}\left|\Wh (\vev\Phi)\right|\
=\ \vev{\frac{1}{3} \Kh_{i\jbar}F^i \Fbar ^{\jbar} }^{1/2} \,\,.
\ee

We note that supersymmetry is also generally broken at any other period of cosmic history when the vacuum energy is non-zero, such as during inflation. We will assume that such spontaneous breaking is also performed by moduli $F$-terms, keeping the details for the next Section.

At this stage, the effective Lagrangian of the observable sector can be written down in a straightforward manner. The  Lagrangian of the effective theory for
$Q^{\alpha}$ and $\Phi^i$ is first written down, and the dynamical moduli fields, including the auxiliary $F$-terms, are replaced by their vevs. The flat limit $M_{pl} \rightarrow \infty$ while keeping $m_{3/2}$ fixed is taken.

We will be especially interested in the potential for the observable scalars (which, by abuse of notation, we call $Q^{\alpha}$). This is given by
\bea \label{effmattpot1}
V_{eff}(Q,\Qbar)  &= &
\sum_{a}{\frac{g_a^2}{4}}
    \left( \Qbar^{\overline{\alpha}} Z_{\overline{\alpha} \beta} T_a Q^{\beta}\right)^2\
+\ \del_{\alpha} W_{eff} Z^{\alpha \overline{\beta}} \bar\del_{\overline{\beta}}\overline{W}_{eff} \nonumber \\
&+& m^2_{\alpha\overline{\beta}}Q^{\alpha} \Qbar^{\overline{\beta}}\
+\ \left(\frac{1}{3} A_{\alpha \beta \gamma}Q^{\alpha} Q^{\beta} Q^{\gamma}\,
    +\,\frac{1}{2} B_{\alpha \beta}Q^{\alpha}Q^{\beta} \ + h.c. \right)
\eea
The first line gives the scalar potential of an effective theory
with unbroken rigid supersymmetry. The second line encodes the soft terms. The soft terms are given in terms of moduli vevs and $F$-terms as follows:
\bea \label{masses11}
m^2_{\alpha \overline{\beta}} \ & =\ m^2_{3/2} Z_{\alpha \overline{\beta}} \ -\ F^i \Fbar^{\jbar} R_{i\jbar \alpha \overline{\beta}} \ + \ V_0 \nonumber \\
A_{\alpha \beta \gamma}\ & = \ F^i D_i Y_{\alpha \beta \gamma} \nonumber \\
B_{\alpha \beta}\ & = \ F^i D_i \mu_{\alpha \beta}\ -\ m_{3/2} \mu_{\alpha \beta}
\eea
where
\bea
R_{i\jbar \alpha \overline{\beta} }\ &
=\ \del_i \bar{\del}_{\jbar} Z_{\alpha \overline{\beta}}\
    -\ \Gamma_{i \alpha}^\gamma Z_{\gamma \bar{\delta}}\overline{\Gamma}^{\bar{\delta}}_{\jbar \overline{\beta}}\,,
    \qquad \Gamma_{i \alpha}^\gamma \ = \ Z^{\gamma \overline{\beta}} \del_i Z_{\overline{\beta} \alpha} \\
D_i Y_{\alpha \beta \gamma}\ &
=\ \del_i \Y_{\alpha \beta \gamma}\ +\ \frac{1}{2} \Kh_i Y_{\alpha \beta \gamma}\
    -\ \Gamma_{i(\alpha}^\delta Y_{\beta \gamma )\delta}^{} \\
D_i \mu_{\alpha \beta}\ &
=\ \del_i \mu_{\alpha \beta}\ +\ \frac{1}{2} \Kh_i \ \mu_{\alpha \beta}\
    -\ \Gamma_{i(\alpha}^\gamma\ \mu_{\beta)\gamma}^{} 
\eea

All quantities appearing in
~\ref{masses11}  are  covariant with respect to the supersymmetric
reparametrization of matter and moduli fields
as well as  covariant under Kahler transformations.

\section{Cosmological Phenomena and Geometric Constraints} \label{cosmo1}

Having described the setting of effective supergravity, we now turn to a discussion of the three cosmological applications of \ref{effV1} and \ref{effmattpot1} mentioned in the introduction. As a unifying theme, we first show how \ref{basicntro1} emerges as a crucial quantity.

\subsection{A Bound on $V^{\prime \prime}(\Phi)$} \label{cosmo1a}

The starting point is to consider the mass matrix
\be \label{Hessian}
N \, = \,\left( \begin{array}{cc}
\nabla^i \nabla_j V  &  \nabla^i \nabla_{\overline{j}} V  \\
\nabla^{\overline{i}}\nabla_j V & \nabla^{\overline{i}} \nabla_{\overline{j}}V   
\end{array} \right) \,\,, \,\,\,\,\,\,\,\, V \, \equiv \, V_{eff}(\Phi,\overline{\Phi}) \,\,,
\ee
where as usual %
\be
\nabla_{i}f^k \equiv \partial_{i}f^k + \Gamma_{ij}^k f^j  
\ee
for any vector $f^k$.

The lightest modulus mass will be denoted by $m^2_{\Phi, {\rm lightest}}$ and is given by
\be
m^2_{\Phi, {\rm lightest}} \, = \, {\rm min \,\,\, eigenvalue} \, \{ N \} \,\,.
\ee

Since $m^2_{\Phi, {\rm lightest}}$ is defined as the minimum eigenvalue of the matrix $N$, it always satisfies a bound. For any given unit vector $u^I$ one has
\be
m^2_{\Phi, {\rm lightest}} \, \leq \, u_I N^I_J u^J \,\,.
\ee

Choosing $u^I$ cleverly, it is possible to obtain simple expressions. The obvious choice is the "preferred" direction in moduli space, along the SUSY breaking direction 
\be
u_I = (e^{-\iota \phi} f_{i},e^{\iota \phi} f_{\ibar})/(\sqrt{2})\,\,,
\ee
where $i = 1 \ldots p$ denote all the SUSY breaking moduli, $\phi$ is an arbitrary phase, and the $f_i$ are aligned along the SUSY breaking directions
\be
f_i = F_{i}/|F|\,\,.
\ee

Taking $u^J = (e^{\iota \phi} f^{i},e^{-\iota \phi} f^{\ibar})/(\sqrt{2})$, one obtains
\be \label{bound1}
m^2_{\Phi, {\rm lightest}} \, \leq \, \nabla_i \nabla_{\jbar} V f^{i}f^{\jbar} \, + \, {\rm Re}(e^{2 \iota \phi} \nabla_i \nabla_j V f^i f^j) \,\,.
\ee
The second piece in \ref{bound1} is superpotential dependent and needs to be eliminated. Choosing $\phi = 0$ and $\phi = \pi/2$ and adding the two resulting versions of \ref{bound1} achieves this and one obtains
\be \label{bound2}
m^2_{\Phi, {\rm lightest}} \, \leq \, \nabla_i \nabla_{\jbar} V f^{i}f^{\jbar}  \,\,.
\ee

It now remains to evaluate the right hand side of \ref{bound2}. One obtains

%
\bea \label{bound3}
m^2_{\Phi, {\rm lightest}} \,\,\,\, &  \leq  \,\,\,\, & 2m^2_{3/2} - (V + 3 m^2_{3/2})R_{i \jbar k \lbar} f^{i} f^{\jbar} f^{k} f^{\lbar}  \nonumber \\
& + & \frac{1}{V + 3 m^2_{3/2}}\nabla_i V \nabla^i V + 4 \left( \frac{m^2_{3/2}}{V + 3m^2_{3/2}} \right)^{\frac{1}{2}}{\rm Re}(\nabla_i V f^i)
\eea
%

We now apply \ref{bound3} to different physical contexts.

\subsection{Slow Roll Modular Inflation} \label{cosmo1b}

For slow roll modular inflation, we note that the slow-roll parameter $\eta$ is given by
\be
\eta = \frac{1}{3 H^2 M^2_{pl}} m^2_{\Phi, {\rm lightest}} \,\,.
\ee
In \ref{bound3}, we drop all terms involving $\nabla_i V \, \sim \, \sqrt{\epsilon}$, since $\sqrt{\epsilon} < \mathcal{O}(10^{-3})$. We also set $V = 3 H^2 M^2_{pl}$ . The spectral index is given by
\be
n_s \, = \, 1+2\eta \,\, \Rightarrow \,\, \eta_{\rm observed} \sim -0.01\,\,\,\,.
\ee
Putting this value on the right hand side of \ref{bound3}, we obtain
\be \label{Rboundinflation}
R[f] \,\,\, \leq  \,\,\, \frac{2}{3} \frac{m^2_{3/2}}{m^2_{3/2} + H^2} 
\ee
The quantity \ref{basicntro1} is thus bounded from the requirement of the flatness of the inflaton potential required to generate sufficient number of e-foldings. While the exact value of the bound depends on the relative values of the gravitino mass and $H$, there is a hard bound:
%
%
\be \label{conditioninflation}
R[f] \, < \,  \frac{2}{3} \,\,.
\ee
%

Several comments are in order. Clearly, the condition is not sufficient - obtaining slow-roll parameters along some modular direction that can lead to acceptable inflationary observables requires full knowledge of the potential and higher order corrections. This will entail knowing the superpotential, for example. However, given a set of Kahler geometries, and asked which ones can, \textit{in principle}, admit modular inflation, the necessary condition \ref{conditioninflation} is most useful as a first check.


\subsection{Thermal History} \label{cosmo1b}

There is another set of applications that one can get from \ref{bound3}, which also involves the quantity \ref{basicntro1}. Since \ref{bound3} is a bound on the lightest modulus, this has implications for the thermal history of the Universe.

The coherent oscillations of a modulus $\Phi$ about its low energy minimum lead to the formation of a scalar condensate, which scales like matter and dilutes more slowly that the primordial radiation produced during reheating. Depending on the initial displacement of the modulus, its energy can come to dominate the energy density of the Universe. Moreover, because it is only gravitationally coupled to other fields, its decay rate is
\be \label{decayrate} 
\Gamma_\Phi = \frac{c}{2\pi} \frac{m_\Phi^3}{\Lambda^2},
\ee
where we expect $\Lambda \sim M_{pl}$ and $c$ depends on the precise coupling in the fundamental Lagrangian, but typically takes values of at most ${\mathcal O}(100)$. Light Standard Model particles that are produced during this decay will `reheat' the universe for a second time.  The corresponding reheat temperature is given by $T_r \sim g_*^{-1/4} \sqrt{\Gamma_\Phi \, M_{pl}} $ or
\be
 \label{Tr}
T_{\rm r} = c^{1/2} \left(\frac{10.75}{g_*}\right)^{1/4} \left( \frac{m_{\Phi}}{50\, {\rm TeV}}\right)^{3/2}\, T_{\rm BBN} \,, 
\ee
where $T_{\rm BBN} \simeq 3 ~ {\rm MeV}$ and $g_*$ is the number of relativistic degrees of freedom at $T_{\rm r}$. The reheat temperature must be larger than around $3$ MeV to be in agreement with light element abundances as predicted successfully by Big Bang Nucleosynthesis. \cite{Kawasaki:1999na}. 

An interesting departure from a thermal post-inflationary universe occurs if the mass of the lightest modulus is in a window between $\mathcal{O}(10) - \mathcal{O}(1000)$ TeV.  For masses that are much higher, little departure from a thermal universe is expected, whereas the lower bound comes from consistency with BBN as discussed before. 

The key question is -- {\it should one generally expect a modulus in this mass range?} To answer this question, we start from \ref{bound3} and set 
\be
\nabla_i V \,\, = \,\, V \,\, = \,\, 0 \,\,.
\ee
This yields the following bound
\be \label{nonthermalbound}
m^2_{\Phi, {\rm lightest}} \, \leq \, 3 m^2_{3/2} \left(\frac{2}{3} - R[f] \right ) \,\,.
\ee

Low energy supersymmetry with gravity mediation typically has the gravitino mass $\mathcal{O}(10) - \mathcal{O}(1000)$ TeV. Given that, non-thermal histories are obtained when \cite{Acharya:2010af}
\be
R[f] \, \sim \, \mathcal{O}(1) \,\,.
\ee

\subsection{Baryogenesis} \label{cosmo1c}

In this subsection, we discuss the connection between Hubble-induced masses and the quantity in \ref{basicntro2}, applying it to the case of Affleck-Dine baryogenesis.

The vacuum energy $V_0$ during inflation breaks supersymmetry. The Affleck-Dine baryogenesis mechanism relies on this supersymmetry breaking to induce tachyonic soft masses along a supersymmetric flat direction. If the flat direction, which we denote by $Q$, is initially displaced from its true minimum, it subsequently oscillates when $V_0$ becomes smaller than the effective mass which is $\sim m_{3/2}$. Depending on the magnitude of the baryon number violating terms in $V(Q)$, a net baryon asymmetry may be produced from the resulting condensate.

The potential for the flat direction $Q$ may be written as
\be
V(Q) = (m_{\rm soft,inf}^2 + m_{\rm soft,final}^2)|Q|^2 + \left( \frac{(A+a_{\rm inf})\lambda Q ^n}{n M_P^{n-3}} + {\rm h.c.} \right) 
+ |\lambda|^2 \frac{|Q | ^{2n-2}}{M_P^{2n-6}} \,\, .
\ee
Here, $m_{\rm soft,inf}^2$ and $a_{\rm inf}$ denote soft parameters induced by supersymmetry breaking during inflation, while $m_{\rm soft,final}$ and $A$ arise from supersymmetry breaking in the final vacuum of the theory. The last term comes from non-renormalizable superpotential contributions.

If $m_{\rm soft,inf}$ is tachyonic, the field $Q$ acquires a non-zero vacuum expectation value during inflation and tracks an instantaneous minimum thereafter, until $H \sim m_{3/2}$. At this point, the field begins to oscillate around the new minimum $Q = 0$ and the soft $A$-term becomes comparable to the Hubble-induced  $a_{\rm inf}$. The field acquires an angular motion to settle into a new phase and the baryon number violation becomes maximal at this time.

From \ref{basicntro2} and \ref{masses11}, and using $F^2 = V_0 + 3 m^2_{3/2}$, the soft masses can be written as
\be
m_{\rm soft,inf}^2  =  V_0 \left(1 - R[f,Q] \right)  +  3m_{3/2}^2 \left(\frac{1}{3} -R[f,Q] \right)  
\ee
Requiring this to be tachyonic, and making the assumption that during inflation $V_0 \, \gg \, m^2_{3/2}$ we thus obtain the result that Affleck-Dine baryogenesis is only possible if 
\be
R[f,Q] \, > \, 1 \,\,.
\ee

\subsection{Summary} \label{cosmo1d}

Summarising the results of the three cases above in terms of $R[f]$ and $R[f,Q]$,  we get the following table.

\begin{table}[!htp] 
\begin{center}
\begin{tabular}{c c c} \hline \hline
\\
Phenomenon          & \,\,\, $R[f]$ \,\,\,  & \,\,\,$R[f,Q]$ \,\,\,   \\ \\ \hline \hline

\\
Modular Inflation           &   $< \, \frac{2}{3}$   & -                    \\ \\
Non-thermal History       &   $\mathcal{O}(1)$   & -                   \\ \\
A-D Baryogenesis    &  -                    &  $> \, 1$             \\

\\ \hline \hline
\end{tabular}
\end{center}
\caption{Conditions on the quantities $R[f]$ and $R[f,Q]$ for the various cosmological phenomena described in the text.}
\label{HoloBiholoresults}
\end{table}

\section{Cosmology and Invariants of Algebraic Forms} \label{invforms1}

In the previous sections, we have described a particularly simple order parameter for a host of cosmological phenomena:
\be
R_{i \jbar k \lbar}f^i f^{\jbar}f^{k}f^{\lbar} \,\,,
\ee
where the indices are summed over the directions in field space along which supersymmetry is broken. We note several features of this order parameter:

\begin{itemize}

\item It depends on the Kahler potential of the theory, since the curvature tensor is derived from the Kahler potential.

\item The expression is field-dependent, so it depends on the allowed values of the moduli and hence implicitly on the superpotential data also.

\item The expression requires knowledge about the SUSY breaking mechanism, to identify the vectors $f^i$ in moduli space along which SUSY is dominantly broken.

\item The expression only makes sense after the correct Kahler coordinates (in which the effective potential takes the form \ref{effV1}) are identified. This is a non-trivial task.

\end{itemize}

The goal is to obtain equivalent conditions on a parameter $\Delta$ that is

\begin{itemize}

\item field-independent

\item independent of knowledge of the SUSY breaking mechanism (hence independent of the orientation of the $f^i$).

\end{itemize}

We will see that this leads naturally into classical algebraic invariant theory. 

To fix the class of Kahler potentials for our effective supergravity, we take the setting of type IIB string theory. We give a full description of the Kahler potential and the identification of Kahler coordinates in Appendix A. The final result is that the Kahler potential is given by the logarithm of the volume form (which is a cubic in the Kahler coordinates $\tau$ that correspond to volumes of four-cycles in the compactified Calabi-Yau):
\beq 
K \,= \,  - 2 \, \ln \mathcal{V}(\tau) \,\,.
\eeq
The volume $\mathcal{V}$ is given by \ref{intnumbers} in terms of the intersection numbers $d^{abc}$, which are given in \ref{intnumbers} as well. For completeness, we collect the expressions here:
\bea\label{intnumbersmaintxt}
  \tau^{a}   &=&  \frac{1}{16} d^{abc} v_b v_c \nonumber \\
  \mathcal{V}  &=& \frac{1}{48} d^{abc}v_a v_b v_c \,\,,
\eea
We note that the $v_a$ denote volumes of two cycles, a basis in which the intersection numbers are naturally expressed. However, they do not constitute the correct Kahler coordinates for the low-energy action, which are provided by the $\tau^a$ that are defined through the Legendre transform in \ref{intnumbersmaintxt}. While obtaining the $\tau^a$ coordinates explicitly starting out from the $v^a$ is hard, in practice one can avoid the problem by working implicitly with \ref{intnumbersmaintxt}. This is shown in Appendix B, where finally the Riemannian curvature has been computed. For easy reference, we display the expression below
\bea \label{riemcurvexp}
R_{i j m n} &=& - g_{im} g_{jn} + e^{-2K} \big(\tilde{d}_{i j k} g^{kl} \tilde{d}_{l m n} 
+ \tilde{d}_{i n k} g^{kl} \tilde{d}_{l j m} \big)  + g_{in} K_j K_m + g_{jm} K_i K_n  \nn\\
&\;&+\, g_{im} K_j K_n + g_{jn} K_i K_m + g_{ij} K_m K_n + g_{mn} K_i K_j - 3 K_i K_j K_m K_n \nn \\
&\;& -\, e^{-K} \big(\tilde{d}_{imj} K_n + \tilde{d}_{imn} K_j + \tilde{d}_{inj} K_m +
\tilde{d}_{nmj} K_i \big)
\eea
We now proceed to compute the quantity $R_{i \jbar m \nbar}f^i f^{\jbar}f^{m}f^{\nbar}$. To this end, we will find it particularly useful to decompose the vectors $f^i$ into directions along $K^i$ and directions $K^i_{\perp}$ that are orthogonal to it. Denoting the unit vectors along those two directions by $k^i$ and $k^i_{\perp}$ respectively, we can write
\be
f^i \,\,\, = \,\,\, \sin{\theta} k^i \, + \, \cos{\theta} k^i_{\perp} \,\,. 
\ee
We note that $k^i_{\perp}$ is itself a vector in a $h^{(1,1)}-1$ dimensional space, parametrized by $h^{(1,1)} - 2$ angles which we can denote by $\theta_{\perp, p}$ with $p = 1 \ldots h^{(1,1)} - 2$.

This is clearly a good strategy, given the structure of \ref{riemcurvexp}. Using moreover the no-scale property $K_i K^i = 3$, the expression reduces to \cite{Covi:2008ea}
\be \label{AiandB}
\frac{2}{3} \, - \, R_{i \jbar m \nbar}f^i f^{\jbar}f^{m}f^{\nbar} \,\, =  \,\, (- A^iA_i \, + \, B) \,\,,
\ee
where $A^i$ and $B$ are functions of the angle $(\theta, \theta_{\perp, p})$, the intersection numbers, and the metric. The full forms of these functions are displayed in Appendix C.

Clearly, it is essential to compute the quantity $A^iA_i + B$, which serves as an order parameter for inflation. In fact, we have
\bea
(- A^iA_i \, + \, B)_{max} & > & 0 \,\, \Rightarrow \,\,\,\, {\rm inflation} \,\, {\rm allowed} \nonumber \\
(- A^iA_i \, + \, B)_{max} & < &  0 \,\, \Rightarrow \,\,\,\, {\rm inflation} \,\, {\rm not \,\, allowed} 
\eea
For a given geometry specified by the intersection numbers, the maximization has to be carried out with respect to the angles $(\theta, \theta_{\perp, p})$ that have been defined earlier.

This is a non-trivial computation, involving a set of coupled cubic equations in $\tan{(\theta, \theta_{\perp, p})}$. In the simplest case of two moduli $h^{(1,1)} = 2$, it can be carried out explicitly, with the result that
\be \label{ABexpr1}
\left(\frac{2}{3} \, - \, R_{i \jbar m \nbar}f^i f^{\jbar}f^{m}f^{\nbar}\right)_{max} \,\, = \,\,k \times \frac{\Delta_2}{24} \, \frac{(\det g)^3}{e^{4 K}}  \,\,  \le  \,\, 1 \ ,
\ee
where the prefactor $k$ is positive, and the entire right hand side of the above equation can be shown to be less than one. For a proof of \ref{ABexpr1}, we refer to Appendix C.

The expression $\Delta_2$ is the discriminant of the volume form (which, for two moduli, is a binary cubic $d^{ijk}v_iv_jv_k$), and is given by
\be \label{discrimquad}
\Delta_2 = -27 \left[(d^{000})^2(d^{111})^2 - 3 (d^{001})^2(d^{011})^2 + 4(d^{000})(d^{011})^3 + 4(d^{001})^3(d^{111})  - 6(d^{000})(d^{001})(d^{011})(d^{111})  \right] \,\,.
\ee
We note that the subscript in $\Delta_2$ signifies that it is the discriminant of a binary cubic; for a general model with $n$ moduli, $h^{(1,1)} = n$, we will be concerned with the discriminant of an $n$-ary cubic, which we will denote by $\Delta_n$.

This has two immediate consequences:

$(i)$ for models with $h^{(1,1)} = 2$, the necessary condition for slow roll modular inflation can be stated as a condition on the discriminant of the volume form:
\bea \label{delta2ordparam}
\Delta_2  & > & 0 \,\, \Rightarrow \,\,\,\, {\rm inflation} \,\, {\rm allowed} \nonumber \\
\Delta_2  & < &  0 \,\, \Rightarrow \,\,\,\, {\rm inflation} \,\, {\rm not \,\, allowed} 
\eea

and

$(ii)$ for models with $h^{(1,1)} = 2$, the canonically normalized moduli masses are bounded by the gravitino mass, from \ref{nonthermalbound}:
\be \label{modboundfor2modcomp}
{\rm moduli \,\, mass \,\, bound:\,\,} m^2_{\Phi, {\rm lightest}} \, \leq \, 3 m^2_{3/2} \,\,.
\ee

\subsection{Algebraic Invariants and a Generalization to $h^{(1,1)} = n$}

The emergence of the discriminant for the case of two moduli is striking. At this point, it is useful to recall some basic facts about classical algebraic invariant theory \cite{Hilbert}. The invariant of a binary form of degree $d$ (by definition in $2$ variables, which, for us, are the moduli)  is a polynomial in the coefficients (which, for us, are the intersection numbers) that remains invariant under the action of the special linear group acting on the variables.

Specifically, we can consider the following binary form of degree $d$
\be
f(x,y) \, = \, \sum_{k=0}^d \binom{d}{k} a_kx^ky^{d-k} \,\,.
\ee
A linear change of variables (under the group $GL_2(C)$), which we label $(c_{ij})$ is a transformation of the variables $x$ and $y$, given by
\be
x \, = \, c_{11}\bar{x} + c_{12} \bar{y} \ \ \ y \, = \, c_{21}\bar{x} + c_{22}\bar{y}
\ee
such that the determinant $c_{11}c_{22} - c_{12}c_{21}$ is nonzero. Under the $SL_2(C)$ action, the binary form is transformed into a new form $\bar{f}(\bar{x},\bar{y})$ in the transformed variables $\bar{x}$ and $\bar{y}$, with coefficients $\bar{a}_k$
\be
\bar{f}(\bar{x}, \bar{y}) \, = \, \sum_{k=0}^{d} \binom{d}{k} \bar{a}_k \bar{x}^k \bar{y}^{d-k} \,\,.
\ee
Clearly, the new coefficients $\bar{a}_k$ are polynomials in the original coefficients $a_i$ and the parameters $c_{ij}$.

A \textit{covariant} of the binary form is defined as a nonconstant polynomial $I(a_0,a_1,\ldots , a_d, x, y)$ such that the following identity holds
\be
I(\bar{a}_0,\bar{a}_1,\ldots , \bar{a}_d, \bar{x}, \bar{y}) \, = \, (c_{11}c_{22} - c_{21}c_{12})^g \, I(a_0,a_1,\ldots , a_d, x, y) \,\,,
\ee
where $g$ is a nonnegative integer. The prefactor on the right hand side is one for a special transformation.

A covariant in which the variables $x$ and $y$ do not occur is called an \textit{invariant}. Every invariant of a binary cubic can be written in terms of the discriminant $\Delta_2$, of degree four, defined previously. Moreover, the algebra of covariants for a binary cubic is generated the discriminant $\Delta_2$, the form itself, its Hessian, and a covariant of degree 3.

A $GL(2,C)$ transformation in our case corresponds to a transformation on the basis of divisors $Div(CY)$ of the Calabi-Yau. It is expected that the necessary criterion for inflation should not change under a basis transformation. However, the important point is that the order parameter is not a \textit{covariant}, but even more strongly, an \textit{invariant} that is completely independent of the stabilized values of the moduli, and specified only by the intersection numbers.

Given the above, we can probe a possible generalization to the case of $h^{(1,1)} = n$, i.e., $n$ moduli. There is an immediate problem with this. For a form of degree $d$ in $n$ variables, the discriminant is a homogeneous polynomial of degree $n.(d-1)^{n-1}$. For us, $d=3$ giving
\be
\Delta_n \,\,\, \rightarrow \,\,\, n.(2)^{n-1} \,\,\,\,{\rm degree \,\, polynomial \,\, in} \,\, d^{ijk} \,\,.
\ee
On the other hand, from the definition of $R_{i \jbar k \lbar}f^i f^{\jbar}f^{k}f^{\lbar}$ and the general form of $A_i$ and $B$ given in Appendix C, it is clear that at any given local maximum (with respect to the angular dependence $(\theta, \theta_{\perp, p})$) of $(-A^iA_i + B)$ is a degree four polynomial in the intersection numbers:
\be
(\frac{2}{3} \, - \, R_{i \jbar m \nbar}f^i f^{\jbar}f^{m}f^{\nbar})_{local \,\, max} \,\,\, \rightarrow \,\,\, 4^{th} \,\,\,\,{\rm degree \,\, polynomial \,\, in} \,\, d^{ijk} \,\,.
\ee
In the limit that all the divisors are identified and only two independent cycles remain, the result must reduce to the invariant $\Delta_2$. Starting with a higher order covariant, it is difficult to see how the moduli dependence will drop out. Thus, the natural solution is to start with a higher order \textit{invariant}, like the discriminant, which naturally reduces to its lower dimensional value upon identification of intersection numbers. To match the polynomial degree, we conjecture that a \textit{product} of local maxima (which is a subset of the total number of extrema, but contains the global maximum) in the $n$ moduli case should give the higher order discriminant:
\be \label{discrimnmodcase}
\prod_{a = 1}^{n.2^{n-3}} \left(\frac{2}{3} \, - \, R_{i \jbar m \nbar}f^i f^{\jbar}f^{m}f^{\nbar}\right)_{crit, a^{th}}  \, \propto \, \Delta_n \frac{(\det g)^{3n.2^{n-3}}}{e^{4n.2^{n-3}K}} \,\,\,\,,
\ee

This result is weaker than the $n=2$ case, where $\Delta_2$ was directly serving as an order parameter in \ref{delta2ordparam}. Moreover, we are unable to calculate the sign of the coefficient on the right hand side. We note that a similar conjecture was arrived at in the case of metastable vacua in heterotic string compactifications \cite{Rathlev:2013wfk}.

However, we can still see that 
\bea \label{discrimnmodcase2}
\left(\frac{2}{3} \, - \, R_{i \jbar m \nbar}f^i f^{\jbar}f^{m}f^{\nbar}\right)_{a^{th}}  \,& \sim &\,  (\Delta_n)^{\frac{1}{n.2^{n-3}}} \frac{(\det g)^{3}}{e^{4K}} \,\sim \, \mathcal{O}(1) \nonumber \\
& \Rightarrow & R[f] \, \sim \, \mathcal{O}(1) \,\,.
\eea
Thus, similar to the case of two moduli, one obtains
\be
m^2_{\Phi, {\rm lightest}} \, \lsim \, 3 m^2_{3/2} \,\,.
\ee
It would be very interesting to work out the exact sign in \ref{discrimnmodcase}, as well as advance  a rigourous proof. We leave that for the future \cite{KSfuture}.

\section{Sectional and Bisectional Curvatures} \label{holobiholosec1}

We have seen in the previous sections that the quantities $R[f]$ and $R[f,Q]$ serve as important order parameters for cosmology within the setting of effective supergravity. In particular, the aim was to express these quantities in as general a form as possible. 

In this section, we define these quantities carefully and look for abstract relations between them. $R[f]$ serves as an order parameter for slow roll modular inflation, while $R[f,Q]$ controls soft masses induced by supersymmetry breaking (during inflation, for example) along visible sector fields. Thus, relations between them will constrain Affleck-Dine baryogenesis, which relies critically on Hubble-induced soft terms.




We begin with a careful definition of the quantities $R[f]$ and $R[f,Q]$. We consider a Kahler manifold of complex dimension $n$, with $R$ denoting its Riemannian curvature tensor. At each point $x$ of $M$, $R$ is a quadrilinear map
\be
T_{x}(M) \times T_{x}(M) \times T_{x}(M) \times T_{x}(M) \, \rightarrow \, \mathbb{R} \,\,.
\ee
Here, $T_{x}(M)$ denotes the tangent space at the point $x$ on the manifold $M$, while $\mathbb{R}$ denotes the reals. 

We can now consider a plane $f$ in the tangent space  $T_{x}(M)$, with an orthonormal basis $(X, Y)$.  The sectional curvature is given by a function on the Grassmann bundle of two-planes in the tangent space at $x$. Specifically, for the plane $f$ at the point $x$ on $M$, the function is given by
\be
K[f] = R(X,Y,X,Y) \,\,.
\ee
The sectional curvature depends on the point $x$ where it is evaluated, and the plane $f$ it is defined for, but not on the choice of basis vectors $(X,Y)$. 

We denote the (almost) complex structure of $M$ by $J$. The set of $J$-invariant planes constitutes a holomorphic bundle over $M$ with fibre $P_{n-1}(C)$. The restriction of the sectional curvature to this complex projective bundle is called the holomorphic sectional curvature:
\be
\mathbb{H}[f] \, = \, R(X, J X, X, J X) \, = \, - \frac{R_{X \overline{X} X \overline{X}}}{g_{X \overline{X}} g_{X \overline{X}}}.
\ee
The bisectional curvature is defined similarly. For two $J$-invariant planes $f$ (with unit vector $X$) and $\mathcal{Q}$ (with unit vector $Q$) in $T_{x}(M)$, the holomorphic bisectional curvature is given by
\be
\mathbb{H}[f,Q] \, = \, R(X, J X, Q, JQ) \, = \, - \frac{R_{X \overline{X} Q \overline{Q}}}{g_{X \overline{X}} g_{Q \overline{Q}}}.
\ee
Like the sectional curvature, this quantity too is independent of the particular choice of basis vectors. Moreover, one trivially has
\be
\mathbb{H}[f,f] \, = \, \mathbb{H}[f] \,\,.
\ee

Moreover, in terms of the quantities $R[f]$ and $R[f,Q]$ that were previously defined, one has
\bea
\mathbb{H}[f] &=& - R[f] \nonumber \\
\mathbb{H}[f,Q] &=& - R[f,Q]
\eea
%


We will be particularly interested in relations between the holomorphic sectional and bisectional curvature. 
%
%

%
At a given point $x$ in the manifold, for orthonormal directions $X$ and $Q$, the holomorphic bisectional curvature $\mathbb{H}[f,Q]$ between the planes $(Q,\Qbar)$ and $(X,\overline{X})$  is a linear combination of holomorphic sectional curvatures of certain planes:
\be \label{associatedrelation}
\mathbb{H}[f,Q] \, = \, \frac{1}{4}\{\sum_{a=1}^{4}\mathbb{H}[\lambda_a] -\mathbb{H}[f] - \mathbb{H}[Q]\}\,\,,
\ee
where the $\lambda_a$ denote certain holomorphic and anti-holomorphic sections associated with the section spanned by the pair $(Q, X)$. For the special case where the holomorphic sectional curvatures are simply constant for all choices of planes in tangent space at $x$ 
\be
R_{j\jbar j \jbar} = {\rm constant} \,\,\, \forall \,\,\, [{\rm span}(\partial_j,\partial_{\jbar}) \in T_{x}(\mathcal{M})] \,\,.
\ee
one obtains
\bea \label{constrelation}
&&\mathbb{H}[f] = {\rm const. \,\, } (c) \nonumber \\
&\Rightarrow & \,\, \frac{|c|}{2} \, \leq \, |\mathbb{H}[f, Q]| \, \leq \, |c| \,\,.
\eea
For orthonormal planes, the lower bound is exactly satisfied. 

We note that a manifold of this type is called a complex space form. If, in addition, the isotropy of the holomorphic sectional curvature in tangent space holds for all $x$ belonging to the Kahler manifold, we say that the manifold has constant holomorphic sectional curvature, of which a maximally symmetric coset space is an example. This is a statement about special components of the curvature tensor; namely, a manifold has constant holomorphic sectional curvature when
\be
R_{j\jbar j \jbar} = {\rm constant} \,\,\, \forall \,\,\, [x \in \mathcal{M}, {\rm span}(j\jbar) \in T_{x}(\mathcal{M})] \,\,.
\ee

Taking the scale of inflation to be high, the conditions for inflation and baryogenesis are
\bea \label{ADandInf}
\mathbb{H}[f]  & > & 0 \nonumber \\
\mathbb{H}[f,Q] &\lesssim & -1 \,\,.
\eea

There is a clear contradiction and we thus have the following no-go result: accommodating \textit{both} slow roll modular inflation and Affleck-Dine baryogenesis is impossible on complex space forms. This is trivial to check explicitly in the special example of maximally symmetric coset spaces.

\section*{Conclusions}

In this paper, we have attempted to construct a universal order parameter within effective supergravity for slow roll modular inflation, non-thermal cosmological histories, and Affleck-Dine baryogenesis. Our starting point was the fact that the local curvature properties of the Kahler manifold spanned by scalars belonging to chiral superfields play a vital role in determining the viability of these diverse phenomena. The Riemannian curvature tensor, evaluated along supersymmetry breaking directions, must necessarily take certain values, summarized in \ref{HoloBiholoresults}.

Next, we have attempted to recast the conditions on the Riemannian curvature in terms of equivalent conditions on other parameters that are entirely field-independent, constructed from the input parameters that specify the Kahler geometry.  For type IIB Calabi-Yau compactifications, in the case of two moduli, we have proven that the order parameter is an invariant (the discriminant) of the Calabi-Yau volume form, neglecting $\alpha^{\prime}$ and $g_s$ corrections. Generalising to the case of $n$ moduli, we have conjectured that the relevant order parameter is the discriminant of a $n-$ary cubic, generated by its ring of invariants. 

We have utilized these results in the case of two moduli to make definitive statements regarding the possibility of slow roll modular inflation depending on the sign of $\Delta_2$. These results are given in \ref{delta2ordparam}. In the case of $n$ moduli we conjecture that it is the sign of $\Delta_n$ that is important, and give partial results towards this conjecture  in \ref{discrimnmodcase}. In the case of two moduli, we are also able to directly prove that the lightest modulus mass is bounded by three times the gravitino mass.

The results in this paper may be useful to rule out certain cosmological phenomena on classes of manifolds based solely on the geometric input that specifies the Kahler potential. As such, they may be useful for computational studies along the lines of \cite{Altman:2014bfa, He:2013epn, Gray:2012jy, Kreuzer:2002uu}.

\section*{Acknowledgements}

We would like to thank Yang-Hui He, Brent Nelson, Scott Watson, and especially Daniel Farquet for illuminating discussions. This work is supported by NASA Astrophysics Theory Grant NNH12ZDA001N.

\section*{Appendix A: Moduli Space and Kahler Coordinates in Type IIB String Theory}

Our goal in this appendix is to provide details for the emergence of the coordinates $\tau^i$, which parametrize the volumes of four-cycles in the internal Calabi-Yau, as the correct Kahler coordinates in type IIB string theory. This provides the background for the assumptions in Section \ref{invforms1}. Along the way, we describe the moduli space of the theory, preparing the ground for the discussion on non-thermal cosmologies in Section \ref{cosmo1b}.

We closely follow the reviews of \cite{Grana:2005jc, Douglas:2006es}, extracting the main results.

\subsection*{A1: Calabi-Yau Moduli Space}

The forms of a Calabi-Yau moduli space are the following: 

$(i)$ one (constant) harmonic 0-form;

$(ii)$ one (3,0) form and one (0,3) form, labelled $\Omega_{CY}$ and $\bar \Omega_{CY}$, respectively;

$(iii)$ a set of $h^{(1,1)}$ harmonic (1,1) and (2,2)-forms. The cohomology basis of the (1,1) forms (respectively, the (2,2) forms) is denoted by $w_a \,\, (\tilde{w_a})$, with $a=1,..,h^{(1,1)}$ ;

$(iv)$ a set of $h^{(2,1)}$ harmonic (2,1) and (1,2)-forms. The cohomology basis of the (2,1) forms (respectively, the (1,2) forms) is denoted by $\chi_k \,\, (\tilde{\chi_k})$, with $ k=1,..,h^{(2,1)}$;

$(v)$ one (3,3) form, the volume $\mathcal{V}$.

For simplicity, we will also sometimes group the basis cycles as follows: $H^{(0)} \oplus H^{(1,1)}$ with basis $w_A=(1, w_a)$, $A=0,..,h^{(1,1)}$; $H^{(3)}$ with basis $(\alpha_K, \beta^K)$, $K=0,..,h^{(2,1)}$.

The four-dimensional effective action before fluxes or orientifolding corresponds to an $\N=2$ ungauged supergravity theory. The strategy is to expand (six-dimensional) internal space deformations of the various fields in the above basis cycles. The (four-dimensional) coefficients for each term of such an expansion will then correspond to visible spacetime fields. These fields are the moduli.

Denoting the internal space coordinates collectively as $y$, and four dimensional spacetime as $x$, we then have
\bea \label{expNS}
 \phi(x,y) &=& \phi(x) \, , \nn\\
g_{i \bj} (x,y)&=& i v^a(x) (\omega_a)_{i \bj}(y) \ , \qquad 
 g_{i j} (x,y)= i \bar z^{k}(x) 
\left(\frac{(\bar \chi_{k})_{i \bar k \bar l} \, 
\Ox^{\bar k \bar l}\,_j}{|\Ox|^2}
\right) (y) \ , \\ 
 B_2 (x,y)&=&  B_2 (x) + b^a(x) \omega_a(y) \, . \nn
\eea
In the above, $g_{i \jbar}$ denotes the metric. From the Neveu-Schwartz (NS) sector, we thus obtain a total of $2\,(h^{(1,1)}+1)+h^{(2,1)}$ $x$-dependent fields or moduli.

A similar expansion can be carried out for the fields belonging to the Ramond-Ramond (RR) sector, which we display below only for type IIB:
\bea \label{expRRIIB}
C_0(x,y)&=& C_0 (x) \ , \nn \\
C_2(x,y)&=&  C_2 (x) + c^a(x) \omega_a(y) \, , \nn \\
C_4(x,y)&=&  V_1^K(x)\, \alpha_K(y) + \rho_a(x) \tilde \omega^a(y) \, 
\eea

Moreover, the Kahler form $J$ is parametrized as
\be
J \, = \, \sum_{a=1}^{h^{(1,1)}} v^a \omega_a \,\,,
\ee
which endows the $v^a$ with a natual interpretation as volumes of two-cycles.

The type IIB moduli are arranged into $\N = 2$ multiplets. Of the fields shown above, the metric deformations $v^a$ and the deformations $b^a$ get arranged into a hypermultiplet of dimension $h^{(1,1)}$.  Similarly, the moduli $z^k$ go to a $h^{(2,1)}$ dimensional vector multiplet, while the fields $B_2$ and $\phi$ go to a one dimensional tensor multiplet. The four-dimensional metric $g_{\mu \nu}$ (which we have not shown) belongs to a gravity multiplet. All these multiplets are completed by fields coming from the RR sector.

We also display some important quantities that can be obtained from the basis 2-cycles $w^a$, the Kahler form $J$, and the moduli $v^a$. These quantities include the intersection numbers of the geometry $d^{abc}$, the volume form $\mathcal{V}$, and volumes of 4-cycles $\tau^a$:
\bea\label{intnumbers}
  d^{abc} &=& \int \omega^a \wedge \omega^b \wedge \omega^c\ ,  \qquad
\qquad
 \qquad 
  d^{ab}  = \frac{1}{8} \int\omega^a \wedge \omega^b \wedge J 
= \frac{1}{8} d^{abc}v_c \\
  \tau^{a}   &=& \frac{1}{16} \int \omega^a \wedge J \wedge J
= \frac{1}{16} d^{abc} v_b v_c \ , \qquad \ 
  \mathcal{V} = \frac{1}{48} \int J \wedge J \wedge J
 = \frac{1}{48} d^{abc}v_a v_b v_c \ .\nonumber
\eea

The $\N=2$ compactification moduli space is thus given by $\mathcal{M}_h \times \mathcal{M}_v$, where $\mathcal{M}_h$ denotes the hypermultiplet moduli space while $\mathcal{M}_v$ is the vector multiplet moduli space. $\mathcal{M}_h$ is a quaternionic manifold while $\mathcal{M}_v$ is a special K\"ahler manifold. The dilaton field is a  hypermultiplet component. Thus, the geometry of $\mathcal{M}_h$ receives both $\alpha'$ and $g_s$ corrections. $\mathcal{M}_v$, on the other hand, is exact at tree level in both $\alpha'$ and $g_s$. The hypermultiplet moduli space $\mathcal{M}_h$ contains a subspace $\mathcal{M}_h^0$ parameterized by vacuum expectation values of NS fields, with the RR moduli being set to zero. We have displayed this parametrization above. At string tree level the subspace $\mathcal{M}_h^0$ has a special Kahler structure.

We next turn to the reduction of this theory to a $\N = 1$ effective supergravity theory, obtained by orientifolding.

\subsection*{A2: Moduli Space for $\N=1$ Supergravity}

The $\N=1$ theory is obtained by gauging a discrete symmetry of the form $(-1)^{\epsilon F_L}\Omega\sigma$ where $\Omega$ denotes world-sheet parity, $F_L$ is left-moving fermion number and $\epsilon$ takes values $0,1$ depending on the model. Moreover, $\sigma: CY \to CY$ is a holomorphic involution of the Calabi-Yau manifold $CY$ which preserves the holomorphic three-form $\Omega_{CY}$ up to sign $\sigma^* \Omega_{CY} = (-1)^\epsilon \Omega_{CY}.$ The value $\epsilon=1$ corresponds to theories with O3/O7 planes. 

The massless spectrum of $\N=1$ orientifold compactifications is naturally organized in vector and chiral multiplets. For orientifolds with O3/O7 planes, there are $h^{2,1}_-$ chiral multiplets which correspond to the invariant complex structure deformations (denoted above by $z^k$), $h^{1,1}_+$ chiral multiplets that correspond to invariant complexified K\"ahler deformations (formed of $v^a$ and $\rho_a$), and $h^{1,1}_-$ chiral multiplets that parameterize the expectation values of the two-form fields $B_2$ (denoted above by $b^a$) and a similar form $C_2$ coming from the RR sector (denoted above by $c^a$). This field content is displayed in ~\ref{ta:IIBOmult}.

\begin{table}
\begin{center}
\begin{tabular}{|c|c|c||c|c|}
\hline 
 & \multicolumn{2}{|c||}{ O3/O7} & \multicolumn{2}{|c|}{O5/O9} \\
\hline
gravity multiplet & $1$ & $g_{\mu\nu}$ &  $1$ & $g_{\mu\nu}$ \\
\hline
vector multiplets & $h^{(2,1)}_+$ & $V_1^\alpha$ & $h^{(2,1)}_-$ & $V_1^k$ \\
\hline
&  $h^{(2,1)}_-$ & $z^k$ &  $h^{(2,1)}_+$ & $z^\alpha $ \\
\cline{2-5} chiral multiplets &
 $h^{(1,1)}_+$ & $(v^\alpha, \rho_\alpha)$  &
 $h^{(1,1)}_+$ & $(v^\alpha, c^\alpha)$  \\
 \cline{2-5}  &
 $h^{(1,1)}_-$ & $(b^a, c^a)$ &
 $h^{(1,1)}_-$ & $(b^a, \rho_a)$   \\
\cline{2-5}
 & $1$ & $(\phi, C_0)$ &  $1$ & $(\phi, C_2)$ \\
\hline
\end{tabular} 
\caption{\label{ta:IIBOmult} Type IIB moduli arranged in $\N=1$ multiplets for O3/O7 and O5/O9 orientifolds.}
\end{center}
\end{table}

Very importantly for all calculations that follow, the moduli space of the $N=1$ theory is a K\"ahler manifold. For small string coupling and large compactification radius the moduli space is a direct product of the complex structure moduli, complexified K\"ahler moduli and a dilaton-axion factor.

By definition, correct Kahler coordinates are those in which the effective action takes the standard $\N=1$ form:
\beq\label{N=1action}
  S^{(4)}_{\N=1} = -\int_{M_4} \tfrac{1}{2}R * \mathbf{1} +
  K_{I \bar J} DM^I \wedge * D\bar M^{\bar J}  
  + \tfrac{1}{2}\text{Re}f_{\alpha \beta}\ 
  F^{\alpha} \wedge * F^{\beta}  
  + \tfrac{1}{2}\text{Im} f_{\alpha \beta}\ 
  F^{\alpha} \wedge F^{\beta} + V*\mathbf{1}\ . 
\eeq
Here $M^I$ denote the complex scalars
in the chiral multiplets. The potential $V$ is given in terms of the superpotential
$W$ and the D-terms $D_{\alpha}$ by 
\beq\label{N=1pot}
V=
e^K \big( K^{I\bar J} D_I W {D_{\bar J} \bar W}-3|W|^2 \big)
+\tfrac{1}{2}\, 
(\text{Re}\; f)^{-1\ \alpha \beta} D_{\alpha} D_{\beta}
\ ,
\eeq
with the K\"ahler covariant derivatives, defined as
\beq \label{Kahlercov}
D_I W = \partial_I W + W \partial_I K \,.
\eeq

In type IIB, the K\"ahler coordinates depend on what kind of orientifold
projection is performed. For O3/O7 projections, these are
the complex structure moduli $z^k$ and 
\begin{eqnarray}\label{tau}
 \xi &=& C_0+i e^{- \phi}\ , \qquad
  G^a = c^a -\xi b^a\ ,\\
  T^\alpha &=& \tau^{\alpha} + i \rho^\alpha 
  - \frac{i}{2(\xi-\bar \xi)}\,  d^{\alpha b c}G_b (G- \bar G)_c  \  , \nonumber
\end{eqnarray}

where the intersection numbers $d^{\alpha b c}$ 
and $\tau^{\alpha}$ have been defined before.

 The K\"ahler potential is 
%
\beq \label{KIIBO3}
K_{\rm{O3/O7}}=  - 2 \, \ln \mathcal{V}(T, G, \xi) - \ln \Big[-i \int \Ox(z) \wedge \bar \Ox(\bar z) \Big]
- \ln \left[ -i (\xi - \bar \xi) \right] \,\,.
\eeq

\section*{Appendix B: Computing the Riemannian Curvature}


In this Section, we present details of the computation of the Riemannian curvature tensor in Type IIB Kahler coordinates. 

We will denote all derivatives with respect to the Kahler coordinates $T^i$ by a lower index. Thus, for example, 
\be \label{curv1}
\frac{\del}{\del T^i} \, = \, \frac{\del}{\del \bar{T}^i} \, = \, \frac{1}{2} \frac{\del}{\del \tau^i}
\ee

For a Kahler manifold, all geometric data such as the metric, connection, and curvature can be obtained by taking repeated derivatives of the Kahler potential. For the Kahler potential relevant for us, this amounts to derivatives of the volume form. One immediately obtains
\be \label{curv2}
\del_{\tau^i} \mathcal{V} \, \equiv \, \mathcal{V}_i \, = \, \frac{1}{32} d^{jkl}v_jv_k\frac{\del v_l}{\del \tau^i} \, = \, \frac{1}{4} v_j \frac{\del \tau^j}{\del v_l} \frac{\del v_l}{\del \tau^i} \, = \, \frac{1}{4} v_i \,\,.
\ee
and from there, the first derivative of the Kahler potential:
\be \label{curv3}
\del_{\tau^i} K \, \equiv \, K_i \, = \, -2 \frac{\mathcal{V}_i}{\mathcal{V}} \, = \, - \frac{v_i}{2 \mathcal{V}} \,\,.
\ee

Interestingly, we are able to obtain the information without explicitly solving for the $\tau^i$ in terms of the $v^i$. The price we have to pay, however, is that the answer contains both sets of coordinates.

Several useful relations that can be expressed purely in the $\tau^i$ coordinates are:
\bea
\tau^i K_i &=& -3/2 \nonumber \\ 
K^i = -2 \tau^i \nonumber \\
K^i K_i = 3 
\eea
The third relation is especially important, since it clarifies the no-scale structure of the geometry.

We now press forward to a computation of the metric. To this end, we define the following two parameters:
\be \label{curv4}
d^{ij} \equiv \frac {\partial \tau_i}{\partial v^j} \, = \, \frac{1}{8} d^{ijk}v_k \,\, , \qquad 
d_{ij} \equiv \frac {\partial v_i}{\partial \tau^j}    \ .
\ee
We note that the first is just the definition from before, while the second relation with the lowered indices $d_{ij}$ is only formally defined. Obtaining it explicitly involves inverting the Legendre transformation between $\tau^i$ and $v^i$ in \ref{intnumbers}, which is in general difficult.

The Kahler metric and its inverse can now be formally defined given all the above relations:
\begin{eqnarray} \label{curv6}
g_{i j} = \frac{1}{2} \, K_{i} K_{j} - \frac 14 \, e^{K/2} d_{ij} \, ,  \qquad 
g^{i j} = 4 \, \tau^i \tau^j - 4 \, e^{-K/2} d^{ij}\,. 
\end{eqnarray}
We note that while the metric $g_{ij}$ is only formally defined in terms of the quantity $d_{ij}$, the inverse metric $g^{ij}$ is  written explicitly and involves a combination of $\tau^i$ and $v^i$ dependence (the latter coming from $d^{ij}$). The inverse metric can also be recast into an equivalent form using the no-scale property:
\be \label{curv7}
g^{i j} =  e^{-K} d^{i j k} K_{k} + K^{i} K^{j}\ . 
\ee

For later reference, we also define the following quantity
\be \label{curv11}
\tilde{d}_{ijk} \equiv g_{ip} g_{jq} g_{kl} d^{p q l} \ . 
\ee
We note that our intersection numbers $d^{pql}$ are defined with raised indices; the quantity $\tilde{d}_{ijk}$ above is purely formal.

We now go on to a computation of the curvature tensor, for which we need the third and fourth derivatives of the metric. However, as we have seen above, it is the inverse metric $g^{ij}$ that is more amenable to a direct computation. We thus find
\bea \label{curv8}
\partial_k g^{i j} &=& e^{-K}  d^{i j m} g_{m k} 
- (g^{i j} - K^{i} K^{j}) K_{k} - \delta^{i}_{k} K^{j} -
\delta^{j}_{k} K^{i} \,,  
\nn\\[1mm]
\partial_{mn} g^{i j}   &=& - e^{- 2K}   d^{i j p} g_{p q} 
d^{q r s}  g_{r m} g_{s n} + \delta^{i}_{m} \delta^{j}_{n}
+ \delta^{i}_{n}\delta^{j}_{m} \,.  
\eea
From here, it is possible to express $K_{ijm}$  and the Riemann tensor as

\bea \label{curv9}
K_{ijm} &=& - g_{i p} (\partial_j g^{p q}) g_{q m} \,, \nn\\ 
R_{i j m n} &=&  - g_{i p} g_{q j} (\partial_{mn}g^{p q}) +  g_{i r} (\partial_m g^{rp}) g_{p q} (\partial_n g^{q s}) g_{s j} \, .
\eea

Inserting the relevant values of the third and fourth derivatives yields \ref{D'Auria:2004cu,Ferrara:1994kg, Covi:2008ea}
\bea \label{curv10}
K_{ijm} &=& e^{-K} \tilde{d}_{ijm} - g_{i j} K_m  - g_{im} K_j - g_{jm} K_i
+ K_i K_j K_{k} \ , 
\nn\\[1mm]
R_{i j m n} &=& - g_{im} g_{jn} + e^{-2K} \big(\tilde{d}_{i j k} g^{kl} \tilde{d}_{l m n} 
+ \tilde{d}_{i n k} g^{kl} \tilde{d}_{l j m} \big)  + g_{in} K_j K_m + g_{jm} K_i K_n  \nn\\
&\;&+\, g_{im} K_j K_n + g_{jn} K_i K_m + g_{ij} K_m K_n + g_{mn} K_i K_j - 3 K_i K_j K_m K_n \nn \\
&\;& -\, e^{-K} \big(\tilde{d}_{imj} K_n + \tilde{d}_{imn} K_j + \tilde{d}_{inj} K_m +
\tilde{d}_{nmj} K_i \big)  \ ,
\eea

\section*{Appendix C: Computing $A_i$ and $B$}

In this Appendix, we will give full expressions for the quantities $A_i$ and $B$ in \ref{AiandB}. Going on, we will compute them explicitly in the case of two moduli, and understand how the discriminant $\Delta_2$ appears in this case, proving \ref{ABexpr1}. We will then give an outline of the full computation in the $n$ moduli case, motivating \ref{discrimnmodcase}.

We first decompose the Goldstino unit vectors in the following way\footnote{We note that one can also define a relative phase $e^{i \delta}$ between the basis vectors $k_i$ and $k^{\perp}_i$. In the case of two moduli, it can be shown that this phase reduces to unity when the function is maximized with respect to $\delta$. While a similar computation is lacking for larger number of moduli, we will assume $e^{i \delta} = 1$ is satisfied at all critical points, for simplicity.}, following the notation of the main text
\be
f_i \,\,\, = \,\,\, \sin{\theta} k_i \, + \, \cos{\theta} k^{\perp}_i \,\,. 
\ee
Using this decomposition directly in the expression for the curvature tensor \ref{curv10}, we obtain
%
%
%
%
%
%
%
%
\be \label{AiandBapp}
\frac{2}{3} \, - \, R_{i \jbar m \nbar}f^i f^{\jbar}f^{m}f^{\nbar} \,\, =  \,\, (- A^iA_i \, + \, B) \,\,,
\ee
where
\bea \label{AiandBfullexprs}
A_{i}  &=&  2 \sqrt{2}\sin{\theta} k^{\perp}_{i} - \frac{1}{\sqrt{2}} e^{-K} P_{ij} d^{j m n} k^{\perp}_{m} k^{\perp}_{n} \nonumber \\
B &=&  \Big(g^{im} g^{jn} - \frac 32 \,e^{-2 K}  d^{i j p} P_{p q} d^{q m n} \Big)
k^{\perp}_{i}  k^{\perp}_{j} k^{\perp}_{m} k^{\perp}_{n}
\label{s2}
\eea

\subsection*{C1: Two Moduli: $h^{(1,1)} = 2$}

 The simplest non-trivial case is that of two moduli, $h^{(1,1)} = 2$. We will see that the main features of the computation are explicit here.

The space perpendicular to $k^i$ is one-dimensional and is spanned by  $k^{\perp i}$. The projection operator appearing in \ref{AiandBfullexprs} is simply given by
\be
P^{ij} = g^{ij} - k^i k^j = k^{\perp i} k^{\perp j} \,.
\ee

The expressions for $A_i$ and $B$ become
\bea \label{AiBin2modcase}
\frac{1}{\cos{\theta}^2}\,  A_i & = & k^{\perp}_i \left[\frac{2\sqrt{2}}{\sqrt{3}}  \tan{\theta} - \frac{1}{\sqrt{2}} e^{-K} d^{p q r} k^{\perp}_p k^{\perp}_q k^{\perp}_r \right] \,\,, \\
\frac{1}{\cos{\theta}^4}\, B &=& \bigg[1 - \frac32 
\Big(e^{-K} d^{p q r} k^{\perp}_p k^{\perp}_q k^{\perp}_r \Big)^2 \bigg] \,\,.
\eea
We will show, in the next subsection, that the quantity appearing in \ref{AiBin2modcase} simplifies to
\bea \label{discr2modemerge}
1 - \frac{3}{2} \Big(e^{-K} d^{p q r} k^{\perp}_p k^{\perp}_q k^{\perp}_r \Big)^2 =  \frac{\Delta_2}{24} \, \frac{(\det g)^3}{e^{4 K}} \,\, (\le \, 1) \, \,.
\eea
The inequality $\le \, 1$ comes from inspection of the left side of the equation.

Putting all of this together, we get
\be \label{seccurv1}
\frac{1}{\cos{\theta}^4}\, (-A^iA_i + B)\, = \,  \left( \frac{\Delta_2}{24} \, \frac{(\det g)^3}{e^{4 K}}\right) - \frac 83 
\left(\tan \theta - \sqrt{\frac{1 -  \left(\frac{\Delta_2}{24} \, \frac{(\det g)^3}{e^{4 K}}\right)}{8}} \right)^2  \,.
\ee
We now need to extremise with respect to $\theta$:
\be \label{thetaextreme}
\partial_{\theta}\, (-A^iA_i + B) \, = \, 0 \,\,.
\ee
Solving \ref{thetaextreme} leads to a cubic equation in $\tan{\theta}$, whose approximate solution leads to the vanishing of the square term in the above equation. One finally obtains
\be
(-A^iA_i + B)_{max} \, \sim \, \frac{64 \left(\frac{\Delta_2}{24} \, \frac{(\det g)^3}{e^{4 K}}\right)} {\left(9 - \left(\frac{\Delta_2}{24} \, \frac{(\det g)^3}{e^{4 K}}\right)\right)^2} \,\,.
\ee
We thus conclude that
\be \label{ABexpr1app}
\left( \frac{2}{3} \, - \, R_{i \jbar m \nbar}f^i f^{\jbar}f^{m}f^{\nbar}\right)_{max} \,\, = \,\,k \times \frac{\Delta_2}{24} \, \frac{(\det g)^3}{e^{4 K}}  \,\,  \le  \,\, 1 \ ,
\ee
which is \ref{ABexpr1}.

\subsection*{C2: Two Moduli: How $\Delta_2$ Emerges}

The purpose of this subsection is to prove \ref{discr2modemerge}. The simplest way to accomplish this is to perform computations in the so-called canonical frame of divisors, and then transform back to the general frame \cite{Gunaydin:1983bi, Cremmer:1984hj, Farquet:2012cs, Brizi:2011jj}. 

The canonical frame is defined as follows. A real invertible matrix $U$ is introduced such that
\be \label{trans1}
\mathfrak v_i \, = \, U_i^j v_j \,\,.
\ee
Simultaneously, the intersection numbers are transformed as
\be \label{trans2}
\mathfrak d^{ijk} \, = \, \alpha (U^{-1})_l^i (U^{-1})_m^j (U^{-1})_n^k d^{lmn} \,\,.
\ee
This leaves the Kahler potential unchanged up to an irrelevant shift 
\be
\mathfrak K \, = \, K - ln \alpha^2 \,\,.
\ee

The transformation $U$ is chosen such that in the canonical frame, the intersection numbers, metric, and two-cycle volumes take the following form
\be
\mathfrak v_i = 2 \sqrt{3} \delta_i^0, \,\,\,\,\,\, \mathfrak g^{ij} = \delta ^{ij}, \,\,\,\,\,\, \mathfrak K = 0 \,\,.
\ee
This is always possible by counting parameters. We note that the transformation $U$ has simply been introduced as a calculational device, taking advantage of the fact that it is a Kahler manifold.

From the above constraints, one gets $K^i = - \sqrt{3} \delta_0^i$, and the intersection numbers in the canonical basis are given by
\be \label{hatdcanN}
\mathfrak d^{000} = \frac{2}{\sqrt{3}}, \,\, \mathfrak d^{00a} = 0, \,\,\mathfrak d^{0ab} = \frac{1}{\sqrt{3}}\delta^{ab}, \,\,\mathfrak d^{abc} = free, \,\,
\ee
with $a,b,c = 1, \ldots , h^{(1,1)}-1$.

The Riemann tensor can be worked out in the canonical frame to be
\bea
R_{0000} &=& \frac{2}{3}, \,\,R_{000a} = 0, \,\,R_{00ab} = \frac{2}{3}\delta_{ab}, \,\,R_{0abc} = \frac{1}{\sqrt{3}}\mathfrak d^{abc}, \nonumber \\
R_{abcd} &=&  -\delta_{ac}\delta_{bd} + \frac{1}{3}\delta_{ab}\delta_{cd} + \frac{1}{3}\delta_{ad}\delta_{bc} + \mathfrak d^{abe} \mathfrak d^{ecd} + \mathfrak d^{ade} \mathfrak d^{ebc} \,\,.
\eea

We now specialize the above equations to the simplest case of two moduli $h^{(1,1)} = n = 2$. Moreover, we have
\bea
\mathfrak k_i  &=& \frac{\mathfrak K_i}{\sqrt{\mathfrak K^i \mathfrak K_i}} \, = \, (-1,0) \nonumber \\
\mathfrak k^{\perp}_i &=&  (0,1) \,\,.
\eea
Thus,
\be
\mathfrak f^i = (\sin{\theta}, \cos{\theta}) \,\,.
\ee

With the simplified expressions, we can easily derive $A_i$ and $B$ in the canonical frame. We obtain
\be \label{seccurv2}
\frac{1}{\cos{\theta}^4}\, (-A^iA_i + B)\, = \,  \left( 1 - \frac{3}{2} \mathfrak{d}^2_{111}\right) - \frac 83 
\left(\tan \theta - \sqrt{\frac{1 -  \left( 1 - \frac{3}{2} \mathfrak{d}^2_{111}\right)}{8}} \right)^2  \,.
\ee
At this point, we note that the discriminant in the canonical frame is easily computed to be
\be
\Delta_{2, can} = 24 - 36 \mathfrak{d}^2_{111}
\ee
We can thus recast \ref{seccurv2} as
\be \label{aibapp3}
\frac{1}{\cos{\theta}^4}\, (-A^iA_i + B)\, = \,  \frac{\Delta_{2, can}}{24} - \frac 83 
\left(\tan \theta - \sqrt{\frac{1 -  \frac{\Delta_{2, can}}{24}}{8}} \right)^2  \,.
\ee

It now remains to transform back to the general frame. From the definition of the canonical frame \ref{trans2}, and using the expression for the discriminant of a general cubic in \ref{discrimquad}, one obtains the following relation between the canonical frame discriminant and the general one \cite{Farquet:2012cs}
\be
\Delta_{2, can} \, = \, \alpha^4 (\det U)^{-6} \Delta_2
\ee
One also has, from the definition of the canonical frame, 
\bea
\mathfrak v^i &=& \alpha (U^{-1})_j^i \tau^j, \,\,\,\,\,\,\, \mathfrak g^{ij} = \alpha^2 (U^{-1})_p^i (U^{-1})_q^j g^{pq}, \nonumber \\
&\Rightarrow & (\det U)^{-6} \, = \, \alpha^{-12}(\det g)^3 \,\,. 
\eea
Since $e^{\mathfrak K} = e^K \alpha^{-2}$, we can combine all of the above to finally get
\be \label{aoeqn2}
\Delta_{2,can} \, = \,  \Delta_2 \,\times \, \frac{(\det g)^3}{e^{4 K}}\,\,.
\ee
Plugging this back to \ref{aibapp3}, we get \ref{ABexpr1}.

\subsection*{C3: $h^{(1,1)} = n$ Moduli Case}

In this subsection, we make some preliminary attempts at solving the case of $h^{(1,1)} = n$ moduli, towards a derivation of \ref{discrimnmodcase}. The strategy is to explore the structure of $A_i$ and $B$, and obtain the generalizations of \ref{AiBin2modcase}, \ref{discr2modemerge}, and \ref{seccurv1}. 

As before, we first decompose the Goldstino direction in components along orthonormal directions spanning the space orthogonal to $K_i$
\bea \label{eq:real_orth_basis}
	K_ik^{\perp i}_{\alpha}=0,\quad k^{\perp i}_{\alpha }k^{\perp i}_{\beta} = \delta_{\alpha\beta},\quad \bar{k^{\perp i}}_\alpha =k^{\perp i}_{\alpha} \quad\text{ for }\, \alpha,\,\beta=1,\dots,\,p-1.
\eea

The projector $P^{ij}$ onto the orthogonal complement of $K^i$ can be written as
\begin{align}
	\label{eq:projector_p}
	P^{ij} = \sum_{\alpha=1}^{p-1}k^{\perp i}_\alpha k^{\perp j}_\alpha .
\end{align}
A general unit vector $K^{\perp i}$ orthogonal to $K^i$ can be parameterized as
\begin{align}
	\label{eq:param_N_p}
	K^{\perp i} = \sum_{\alpha=1}^{p-1} e^{ {\rm Im}\varphi_\alpha}c_\alpha k^{\perp i}_\alpha
\end{align}
with real phases $\varphi_\alpha$ and real $c_\alpha$ satisfying
\begin{align}
	\sum_{\alpha=1}^{p-1} c_\alpha^2=1.
\end{align}

$B$ can now be written as 
\begin{align}
	\label{equationforB}
	B &= 1 - \frac{3}{2} e^{-2K} \sum_{\alpha\beta\gamma\delta\eta}c_\beta c_\gamma c_\delta c_\eta D_{\alpha\beta\gamma}D_{\alpha\delta\eta} \,\,,
\end{align}
where the symmetric rank 3 tensor
\begin{align}
  \label{theDs}
  D_{\alpha\beta\gamma} := d_{ijk}k^{\perp i}_\alpha k^{\perp j}_\beta k^{\perp k}_\gamma
\end{align}

At this point, it is simplest to go to the canonical frame, where something like \ref{seccurv1} can be derived. Specifically, we define 
\be \label{canonbexpr1}
\mathfrak{b}_{abcd}^{\mathcal{O}} \, \equiv \, \left[ \frac{1}{3} \delta^{ab} \delta^{cd} - \frac{1}{2} \mathfrak d^{abe} \mathfrak d^{ecd} \right] + \left[ \frac{1}{3} \delta^{ac} \delta^{bd} - \frac{1}{2} \mathfrak d^{ace} \mathfrak d^{ebd} \right] + \left[ \frac{1}{3} \delta^{ad} \delta^{bc} - \frac{1}{2} \mathfrak d^{ade} \mathfrak d^{ebc} \right]
\ee
and
\be \label{canonbexpr2}
\mathfrak{B} \, \equiv \, \mathfrak{b}_{abcd}^{\mathcal{O}}f^a f^b f^c f^d \,\,.
\ee
Then, in the canonical frame, we can write
\be \label{canoncritexpr1}
\left( \frac{2}{3} \, - \, R_{i \jbar m \nbar}f^i f^{\jbar}f^{m}f^{\nbar}\right) \, = \,  \mathfrak{B} - \frac{8}{3} \sum_{e} \left[f^0 f^e + \frac{\sqrt{3}}{4} f^a f^b \mathfrak d^{abe} \right]^2 \,\,,
\ee
where we have 
\be \label{canonBnmoduli}
\mathfrak{B} \,\, \equiv \,\, \mathfrak{B}(\mathfrak{c}_\alpha) \,\, \equiv \,\, \mathfrak{B}(\theta, \theta_p) \,\,.
\ee
We have displayed the fact that $\mathfrak{B}$ s a function of the expansion coefficients $c_\alpha$ of the Goldstino directions $f^i$, or equivalently of the angles $\theta_p$ that fix the $f^i$. 

In the case of two moduli, the angular dependence was simple and was displayed in \ref{seccurv1}. The maximisation procedure led to the approximate vanishing of the square term. In this case, the angular dependence is more complicated, and a set of coupled cubic equations in $\tan{\theta_p}$ must be solved. Nevertheless, we can conjecture that at least a subset of maxima, including the global one, corresponds to the case when the square term in \ref{canoncritexpr1} vanishes. Thus, we have
\be
\left( \frac{2}{3} \, - \, R_{i \jbar m \nbar}f^i f^{\jbar}f^{m}f^{\nbar}\right)_{local \,\, max} \, = \, \mathfrak{B}_{local \,\, max} \,\,.
\ee

It now remains to determine the structure of $\mathfrak{B}_{local \,\, max}$. It should be clear from \ref{canonbexpr1} and \ref{canonbexpr2}, as well as the case of two moduli, that any given local maximum $\mathfrak{B}_{local \,\, max}$ is a fourth degree polynomial in intersection numbers. Moreover, the factors of $\det{g}$ and $e^K$ work out the same way as the two moduli case, (\ref{cantogen1}, \ref{cantogen2}, and \ref{aoeqn2}), when going from the canonical to the general frame. Thus, we arrive at conjecture \ref{discrimnmodcase}.

\section*{Appendix D: Invariants of Three and Four Moduli}

In the case of three moduli, we have
\be
\prod_{a = 1}^3 \left(\frac{2}{3} \, - \, R_{i \jbar m \nbar}f^i f^{\jbar}f^{m}f^{\nbar}\right)_{crit, a^{th}} \, \propto \, \Delta_3 \frac{(\det g)^9}{e^{12K}}\,\,.
\ee

The ring of invariants of ternary cubics is generated by the Aronhold invariants $S$ and $T$, which are homogeneous polynomials of degree 4 and 6, respectively, in the coefficients of the cubic \cite{aronhold}. The discriminant is a homogeneous polynomial of degree 12, given by
\be
\Delta_3 \, = \, S^3 - T^2 \,\,.
\ee

For convenience, we give the polynomials $S$ and $T$ below. For the cubic $f(x,y,z)$
\be
f(x,y,z) = a x^3 + b y^3 + c z^3 + 3 d x^2 y + 3 e y^2 z +  3 f z^2 x + 3 g x y^2 + 3 h y z^2 + 3 i z x^2 + 6 j x y z
\ee
we have
\bea
S &=& a g e c - a g h^2 - a j b c + a j e h + a f b h - a f e^2 - d^2 e c + d^2 h^2 + d i b c - d i e h + d g j c - d g f h - 2 d j^2 h + \nonumber \\
&+& 3 d j f e - d f^2 b - i^2 b h + i^2 e^2 - i g^2 c + 3 i g j h - i g f e - 2 i j^2 e + i j f b + g^2 f^2 - 2 g j^2 f + j^4
\eea
and
\bea
T &=& a^2 b^2 c^2 - 3 a^2 e^2 h^2 - 6 a^2 b e h c + 4 a^2 b h^3 + 4 a^2 e^3 c - 6 a d g b c^2 + 18 a d g e h c - 12 a d g h^3 + 12 a d j b h c  \nonumber \\
 &-& 24 a d j e^2 c + 12 a d j e h^2 - 12 a d f b h^2 + 6 a d f b e c + 6 a d f e^2 h + 6 a i g b h c - 12 a i g e^2 c + 6 a i g e h^2 + 12 a i j b e c \nonumber \\
&+&  12 a i j e^2 h - 6 a i f b^2 c + 18 a i f b e h - 24 a g^2 j h c - 24 a i j b h^2 - 12 a i f e^3 + 4 a g^3 c^2 -  12 a g^2 f e c + 24 a g^2 f h^2\nonumber \\
 &+& 36 a g j^2 e c + 12 a g j^2 h^2 + 12 a g j f b c - 60 a g j f e h - 12 a g f^2 b h + 24 a g f^2 e^2 - 20 a j^3 b c - 12 a j^3 e h \nonumber \\
&+&  36 a j^2 f b h + 12 a j^2 f e^2 -24 a j f^2 b e + 4 a f^3 b^2 + 4 d^3 b c^2 - 12 d^3 e h c + 8 d^3 h^3 + 24 d^2 i e^2 c - 12 d^2 i e h^2  \nonumber \\
&+&  12 d^2 g j h c + 6 d^2 g f e c - 24 d^2 j^2 h^2 - 12 d^2 i b h c - 3 d^2 g^2 c^2 - 24 g^2 j^2 f^2 + 24 g j^4 f - 12 d^2 g f h^2 + 12 d^2 j^2 e c  \nonumber \\
&-& 24 d^2 j f b c -  27 d^2 f^2 e^2 + 36 d^2 j f e h + 24 d^2 f^2 b h + 24 d i^2 b h^2 - 12 d i^2 b e c - 12 d i^2 e^2 h + 6 d i g^2 h c - 60 d i g j e c \nonumber \\
&+& 36 d i g j h^2 + 18 d i g f b c - 6 d i g f e h + 36 d i j^2 b c - 12 d i j^2 e h - 60 d i j f b h + 36 d i j f e^2 + 6 d i f^2 b e + 12 d g^2 j f c  \nonumber \\
&-& 12 d g j^3 c -  12 d g j^2 f h + 36 d g j f^2 e 12 d g f^3 b + 24 d j^4 h + 12 d j^2 f^2 b + 4 i^3 b^2 c + 24 i^2 g^2 e c - 27 i^2 g^2 h^2  \nonumber \\
&-& 36 d j^3 f e  - 12 i^3 b e h + 8 i^3 e^3 - 24 i^2 g j b c + 36 i^2 g j e h + 6 i^2 g f b h + 12 i^2 j^2 b h - 3 i^2 f^2 b^2 - 12 d g^2 f^2 h   \nonumber \\
 &-& 12 i^2 g f e^2 - 24 i^2 j^2 e^2 + 12 i^2 j f b e - 12 i g^3 f c + 12 i g^2 j^2 c + 36 i g^2 j f h - 12 i g^2 f^2 e - 36 i g j^3 h  \nonumber \\
 &-& 12 i g j^2 f e + 12 i g j f^2 b + 24 i j^4 e - 12 i j^3 f b + 8 g^3 f^3 - 8 j^6
\eea

Similar to the case of three moduli, the ring of invariants for the case of a cubic in four variables is generated by homogeneous polynomials $I_8, I_{16}, I_{24}, I_{32}, I_{40}, I_{100}$, where the subscript denotes the degree of the polynomial in the coefficients of the cubic. The discriminant of a quaternary cubic is given by 
\be
\Delta_4 \, = \, (I_8^2 - 64 I_{16})^2 - 2^{11} (I_8 I_{24} + 8 I_{32}) \,\,.
\ee

For a form of degree $d$ in $b$ variables, the discriminant $\Delta_{d,b}$ is a homogeneous polynomial of degree $b \cdot (d-1)^{b-1}$.

\end{document}